# Abnormal Staebler-Wronski effect of amorphous silicon

Wenzhu Liu[1]*[†], Jianhua Shi[1,2]*, Liping Zhang[1,2]*, Anjun Han[1,2]*, Shenglei Huang[1], Xiaodong Li[1], Jun Peng[3], Yuhao Yang[1], Yajun Gao[4], Jian Yu[5], Kai Jiang[1], Xinbo Yang[6], Zhenfei Li[1], Junlin Du[1,2], Xin Song[7], Youlin Yu[1], Zhixin Ma[1], Yubo Yao[1], Haichuan Zhang[2,4], Lujia Xu[4], Jingxuan Kang[4], Yi Xie[2,8], Hanyuan Liu[2,8], Fanying Meng[1,2], Frédéric Laquai[4], Zengfeng Di[9], Zhengxin Liu[1,2][†]

[1]Research Center for New Energy Technology, Shanghai Institute of Microsystem and Information Technology (SIMIT), Chinese Academy of Sciences (CAS), Shanghai 201800, China

[2]Zhongwei New Energy (Chengdu) Co., Ltd, Sichuan 610200, China

[3]Research School of Electrical, Energy and Materials Engineering, The Australian National University, ACT 2600, Australia

[4]King Abdullah University of Science and Technology (KAUST), KAUST Solar Center (KSC), Thuwal 23955-6900, Saudi Arabia

[5]Institute of Photovoltaics, Southwest Petroleum University, Sichuan 610500, China

[6]College of Energy, Soochow Institute for Energy and Materials Innovations (SIEMIS), Soochow University, Suzhou 215006, China

[7]School of Materials Science and Engineering, Jiangsu Collaborative Innovation Center of Photovoltaic Science and Engineering, Changzhou University, Changzhou 213164, China

[8]Tongwei Solar Co., Ltd, Sichuan 610200, China

[9]State Key Laboratory of Functional Materials for Informatics, Shanghai Institute of Microsystem and Information Technology, Chinese Academy of Sciences, Shanghai 200050, China

*These authors contributed equally to this work.

[†]Correspondence to: wenzhu.liu@mail.sim.ac.cn (W.L.); z.x.liu@mail.sim.ac.cn (Z.L.)

## ABSTRACT

Great achievements in last five years, such as record-efficient amorphous/crystalline silicon heterojunction (SHJ) solar cells and cutting-edge perovskite/SHJ tandem solar cells, place hydrogenated amorphous silicon (a-Si:H) at the forefront of emerging photovoltaics. Due to the extremely low doping efficiency of trivalent boron (B) in amorphous tetravalent silicon, light harvesting of aforementioned devices are limited by their fill factors (FF), which is a direct metric of the charge carrier transport. It is challenging but crucial to develop highly conductive doped a-Si:H for minimizing the FF losses. Here we report intensive light soaking can efficiently boost the dark conductance ($\sigma_{dark}$) of B-doped a-Si:H "thin" films, which is an *abnormal* Staebler-Wronski effect. By implementing this *abnormal* effect to SHJ solar cells, we achieve a certified power conversion efficiency (PCE) of 25.18% (26.05% on designated area) with FF of 85.42% on a 244.63-cm$^2$ wafer. This PCE is one of the highest reported values for total-area "top/rear" contact silicon solar cells. The FF reaches 98.30 per cent of its Shockley-Queisser limit.

Hydrogenated amorphous silicon (a-Si:H) is a technologically important semiconductor for transistors, batteries, hydrogen production and solar cells (*1-4*). It has a long history of use in photovoltaic applications because it offers low defect density and tunable conduction type (*5-7*). These optoelectronic advantages strongly rely on configurations of H and Si in the three-dimensional space, described by the radial distribution function (*8*), so precise control of its microscopic structure (*9-11*) is a critical factor to achieving good devices. As B is a trivalent element, it is challenging to establish four-coordinated B−Si$_4$ compounds in the disordered a-Si:H matrix, reported approaches focusing on eliminating invalid Si$_x$−B−H$_y$ doping configurations (Fig. S1) include optimizing the B$_2$H$_6$ flow rate and post-deposition annealing. However, a lack of understanding of the complicated conductive mechanism of B-doped a-Si:H (*p*-a-Si:H) has prevented to develop the full potential of relevant devices. Here we report intensive light soaking is a "fast" means to improving $\sigma_{dark}$ of *p*-a-Si:H thin films that enables highly efficient SHJ solar cells.

Since 1977, light soaking on "thick" a-Si:H films has been widely studied in the research field of a-Si:H thin-film solar cells, but only a small number of works pay attention to its effect on "thin" a-Si:H films, particularly in the research field of SHJ solar cells. Although a few groups report light soaking improves FF of SHJ solar cells by a magnitude of ~0.7%$_{abs}$ (*12*), the hidden mechanisms are still very unclear. We use *in-situ* methods to monitor the time-dependent changes of *p*-a-Si:H thin films during illuminations. Thin *p*-a-Si:H films are deposited on silica glasses, followed by evaporating silver strips to form the transmission-line-model structure. The *in-situ* current-voltage data (Fig. 1A) shows that $\sigma_{dark}$ of the *p*-a-Si:H thin film steadily increases during one-sun illumination, reaching $\sigma_{dark}/\sigma_{dark0}$ ~ 4.71 after 30 minutes. This phenomenon is strikingly contrast to the light-induced degradation of $\sigma_{dark}$ observed in "thick" intrinsic, *p*-type and *n*-type a-Si:H films (*13-15*), supporting the perspective that accumulated stress in "thick" films plays an important role in the Staebler-Wronski effect (*16*), because the maximum stress is roughly proportional to the thickness of the film. After turning off the illumination, the $\sigma_{dark}$ gradually decays (close) to its initial value after more than 1000 minutes (Fig. 1B). Such a decay behavior well fits to a combination of Debye model and Williams-Watts model (Fig. 1C),

$$\Delta\sigma_{dark}(t) = \Delta\sigma_D \exp\left[-(t/\tau_D)^{\beta_D}\right] + \Delta\sigma_{WW} \exp\left[-(t/\tau_{WW})^{\beta_{WW}}\right], \quad (1)$$

$\Delta\sigma_D$, $\Delta\sigma_{WW}$ and $\tau_D$, $\tau_{WW}$ are constant coefficients and characteristic time constants of the Debye model and the Williams-Watts model respectively. Detailed parameters are summarized in Table S1. The Debye model with $\beta_D = 1$ describes free diffusion, while the Williams-Watts model with $0 < \beta_{WW} < 1$ describes a continuous-time random walk composed of an alternation of "steps" and "pauses" (*17*). Examples of the Williams-Watts model include spin-correlation in Cu-Mn and Ag-Mn, spin-glass transition in $BaFe_{12}O_{19}$, dielectric relaxation in $K_{0.3}MoO_3$, spin-lattice relaxation in $\kappa$-$(ET)_2Cu[N(CN)_2]Br$, specific heat in $Fe_xZr_{1-x}$ and H relaxation in a-Si (*18-23*). The good fitting in Fig. 1C suggests the *abnormal* Staebler-Wronski effect is mediated by two independent mechanisms that respectively control the "fast" Debye relaxation and the "slow" Williams-Watts relaxation.

To determine the implicit mechanisms, we investigate the H configurations in *p*-a-Si:H thin films by time-of-flight secondary ion mass spectrometry (ToF-SIMS). The H⁻ spectra (Fig. 2A) show that 30-minute annealing at 180 °C only slightly changes the H content in intrinsic a-Si:H (*i*-a-Si:H), in contrast, the same annealing process expels at least ~21.3% of the H content out of the *p*-a-Si:H. As shown in Fig. S2, ToF-SIMS spectra also reveal that room-temperature oxidation of an *i*/*p*-a-Si:H stack hardly changes the H content in the *i*-a-Si:H film, however, the same oxidation process getters ~17.1% of the H content in *p*-a-Si:H from inside to surface. Based on these findings, we conclude that the B doping plays a crucial part in the formation of metastable H configurations in *p*-a-Si:H.

Next we combine molecular dynamics and *ab initio* dynamics to search the possible binding configurations of aforementioned metastable H. *Ab initio* dynamics observes displacement of four-coordinated Si atoms by B atoms shortens the bonds from ~2.35Å to ~2.07Å (Fig. S3), well consistent with the results of Pandey *et al* (*24*). Further simulations demonstrate these B−Si$_4$ sites have a large probability to capture H atoms to form metastable B−H−Si configurations when diffusive H atoms pass by (Fig. S4), which is in agreement with the reported NMR signal (*25*) and relevant simulations (*26*). As a consequence, conductance of *p*-a-Si:H is expected to decline due to reduction in the quantity of B−Si$_4$ (*24*). Fortunately, transition-state surveys (Fig. 2B) prove that the hopping barriers of H from B−H−Si and Si−H−Si to adjacent Si−H−Si are merely 0.99 ± 0.22 eV and 0.40 ± 0.11 eV respectively, and the energy required to emit H from Si−H−Si is only ~0.97 ± 0.26 eV. These reactions show noticeably lower barriers than that required to emit H from Si−H or B−H single bonds (Fig. 2C), thus account for why there exists much more metastable H configurations in *p*-a-Si:H than that in *i*-a-Si:H (inferred from Fig. 2A). When intensive light soaking provides photons with energy over ~1.0 eV, huge amounts of Si$_3$−B−H should be converted to B−Si$_4$ via the H hopping illustrated in Fig. 2B, resulting in improvement of $\sigma_{dark}$ as has been confirmed in Fig. 1A. This microscopic dynamics is supported by the evidence of light-induced formation of Si−H−Si configurations from low-temperature infrared spectroscopy (*27*).

This mechanistic understanding is also evident in optoelectronic analysis. Olibet *el al*. demonstrates that effective minority carrier lifetime ($\tau_{eff}$) of passivated c-Si is dominated by field passivation and chemical passivation respectively at low injection level and high injection level (*28*). We prepare symmetric passivated structures of *p*-a-Si:H/*i*-a-Si:H/*n*-c-Si/*i*-a-Si:H/*p*-a-Si:H and *i*-a-Si:H/*n*-c-Si/*i*-a-Si:H, and measure their injection-dependent $\tau_{eff}$ before and after two-hour light soaking under one-sun illumination, as well as that after 15-minute annealing at 180 °C. Interestingly, the right graph in Fig. 2D shows the $\tau_{eff}$ of *p*-a-Si:H/*i*-a-Si:H/*n*-c-Si/*i*-a-Si:H/*p*-a-Si:H increases substantially at low injection level, indicating the field passivation is improved with the greatest contribution from better B doping, also intriguing is that the low-injection-level $\tau_{eff}$ return to their initial values after a 15-minute brief annealing at 180 °C; in contrast, the $\tau_{eff}$ remain unchanged at the high injection level, suggesting the light soaking has negligible impact on the chemical passivation. As a comparison, the left graph in Fig. 2D shows the $\tau_{eff}$ of *i*-a-Si:H/*n*-c-Si/*i*-a-Si:H remains constant after either light soaking or annealing. Moreover, ultrafast and broadband transient absorption (TA) signals (Fig. 2E) indicate the light soaking also increases the mobility of photon-generated carriers from $7.10 \times 10^{-3}$ cm$^2 \cdot$V$^{-1}$ s$^{-1}$ to $1.81 \times 10^{-2}$ cm$^2 \cdot$V$^{-1}$ s$^{-1}$. According to Sinton and other researcher (*29*), pseudo FF (pFF) of silicon solar cells includes the effect of space charge recombination, but excludes the influence of resistance. We fabricate devices with structure of Ag/IWO/*p*-a-Si:H/*i*-a-Si:H/*n*-c-Si/*i*-a-Si:H/*n*-a-Si:H/IWO/Ag, then probe their pFF before and after two-hour light soaking under one-sun illumination, as well as that after 15-minute annealing at 180 ℃. Fig. 2F shows the pFF maintains ~86.4% regardless of the light soaking and annealing. This again definitely demonstrates the light soaking makes no difference to the chemical passivation. Consideration of the $\sigma_{\text{dark}}$ (Fig. 1), $\tau_{\text{eff}}$ (Fig. 2D), carrier mobility (Fig. 2E) and pFF (Fig. 2F) leads to the conclusion that the *abnormal* Staebler-Wronski effect stems from activation of B doping rather than reduction of defects in the *p*-a-Si:H. In this regard, we further ascribe the dark decay of $\sigma_{\text{dark}}$ in Fig. 1B to the detrimental reconstruction of B−H−Si configurations, because binding energy of H in B−H−Si shows ~0.59 eV higher over that in Si−H−Si. In accordance, the "fast" Debye relaxation and the "slow" Williams-Watts relaxation (Fig. 1C) are attributed to incorporation of "fast" diffusive H and "slow" hopping H into the B−Si bonds respectively, forming invalid B doping that negatively effects on the $\sigma_{\text{dark}}$ as has been confirmed in Fig. 1B.

Next we use this *abnormal* Staebler-Wronski effect to improve performance of SHJ solar cells. We fabricate devices featuring a structure showcased in Fig. 3A, where thickness of the *p*-a-Si:H is ~15 nm (Fig. S5), their initial open-circuit voltage ($V_{\text{oc}}$), short-circuit current density ($J_{\text{sc}}$), FF and PCE are 744.30 ± 0.68 mV, 38.43 ± 0.07 mA/cm$^2$, 83.70 ± 0.22 % and 23.94 ± 0.04 % respectively, based on 316 continuous devices from our daily production line. Under one-sun illumination, as expected, FF of these cells exhibits a steady increase (Standard Cell in Fig. 3B). The slope of the current-voltage curve near the low-internal-field region ($V_{\text{oc}}$ condition) serves as an indication of charge collection efficiency (*30*), as found in Fig. S6, the light soaking continuously increases the slope near this low-internal-field region, indicating more efficient charge extraction due to enhancement of the net field across the depletion region. This strongly supports our perspective that the light soaking activates better B doping. In contrast, we observe a noticeable drop in the gain of FF for devices annealed for two hours at 180 ℃ (180 ℃ in Fig. 3B), greatly attributed to its less metastable H configurations (inferred from Fig. 2A). It is also worth noting that, when the cells are applied by 13 amperes current (13A in Fig. 3B), the FF exhibits a quite similar behavior to that under one-sun illumination. This implies the photon energy from light soaking is not the exclusive cause responsible for the *abnormal* Staebler-Wronski effect, photons from radiative recombination caused by photon-generated carriers or current-injected carriers also take effect.

Further augmenting the light intensity from one sun to 11 suns, 48 suns and 60 suns, we boost the FF by $0.32 \pm 0.18$ %$_{abs}$, $0.39 \pm 0.14$ %$_{abs}$, $1.40 \pm 0.26$ %$_{abs}$ and $1.50 \pm 0.37$ %$_{abs}$ respectively (Fig. 3C). This highlights that intensive light soaking activates more efficient B doping by pumping more metastable H from B−H−Si to other configurations, in this consideration, we naturally regard SHJ solar cells as the premium choice for concentrator photovoltaic systems. At mass-production level, 60-sun illumination obtains state-of-the-art industrial FF and PCE of $85.19 \pm 0.18$ % and $24.46 \pm 0.05$ % respectively (Fig. 3D, E), together with improved $V_{oc}$ (Fig. S7), thanks to improvement of the build-in field in c-Si absorber. After capping an 80-nm SiO$_x$ antireflection layer onto a high-efficient cell, we submit it to an independent testing center and achieve a certified PCE of 25.18% (26.05% on designated area) with FF of 85.42% on a 244.63-cm$^2$ wafer (Fig. 3F and Fig. S8). They are amongst the highest certified PCE and FF for total-area "top/rear" contact silicon solar cells (*31*) (Fig. 3G and Fig. S9). The FF reaches 98.30 per cent of its Shockley-Queisser limit ~ 86.9% (*32*). In consideration of stability, FF and PCE of devices retain 98.70% and 97.59% of their initial values after 1000-hour damp-heat impact at 85 ℃ and 85% relative humidity (DH85; Fig. 3H), without any encapsulations. In the module level, Fig. 3H shows FF and PCE retain 98.1% (96.8%) and 95.5% (95.4%) respectively after 3000-hour damp-heat impact at DH85 (600 thermal cycles between -40 ℃ and 85 ℃), demonstrating their high stability against extreme climate degradation factors.

Lastly, we explore reversible behavior of the *abnormal* Staebler-Wronski effect. As found in Fig. 4A, we alternately measure the cells' FF under one-sun illumination for 180 minutes and place them in the dark for 720 minutes. Evidently, the FF decays ~0.3−0.35%$_{abs}$ during each "sleeping" in the dark. From Fig. S10, we find the FF rapidly declines by ~0.15%$_{abs}$ in the first ~20 minutes, followed by a slow decay in the next ~745 minutes. This "fast" decay time ~20 minutes is consistent with the characteristic time constant $\tau_D$ ~15.84 ± 1.55 minutes of the Debye relaxation (Table S1), confirming the decay of FF does results from the doped a-Si:H film. Fig. 4A also reveals the FF rapidly climbs up after turning on the light soaking. Thus, the output of power plants made of SHJ solar cells undergoes rapid advancements after sunrise in sunny days, which challenges the up-to-date IEC testing standards, because the in-house certification underestimates their performance in real operations. The following provides a feasible pathway to freezing the dark decay. We take 198 solar cells from the same batch, and divide them into 11 groups. First, the devices in each group undergo a 70-second light soaking under 60-sun illumination, followed by a 25-minute "sleeping" in the dark to finish the "fast" Debye relaxation. Then their FFs are measured before and after 10-minute annealing at different temperatures, as shown in Fig. 4B, the decay magnitude of FF (from Williams-Watts relaxation) dramatically drops when the temperature is decreased from 200 ℃ to 60 ℃, suggesting the low temperature arrests the unfavorable formation of B−H−Si configurations. This observation agrees with the perspective that annealing can accelerate annihilation of Si−H−Si configurations (*33*). Generally, annealing declines the defect density at *i*-a-Si:H/c-Si interface (*34*), thus the phenomenon in Fig. 4B cannot be ascribed to the changes of chemical passivation, which once again excludes the possibility that the *abnormal* Staebler-Wronski effect originates from improvement of chemical passivation at c-Si surface. Using the average ΔFF (Fig. 4B), we derive the temperature-dependent characteristic time constant $\tau_{WW}$ by the Williams-Watts model,

$$\Delta\text{FF}(t) = \Delta\text{FF}(0)\exp\left[-(t/\tau_{WW})^{\beta_{WW}}\right] \quad . \tag{2}$$

According to Kakalios *et al* (*35*), a-Si:H's $\beta_{WW} = 0.00165T$ (in Kelvin) independent of the doping type. The $\tau_{WW}$ on the other hand obeys an Arrhenius relationship,

$$\ln(1/\tau_{WW}) = \ln(1/\tau_{WW0}) - E_a/(RT) \quad , \tag{3}$$

*R* is molar gas constant. Fig. 4C shows the fitting of equation (3) to the $\tau_{WW}$ (blue circles) calculated from equation (2), interestingly, the theoretical $\tau_{WW}$ (red circle) from Table S1 is close to the extrapolation of the fitting line, confirming the validation of equation (3). The derived activation energy $E_a \sim 0.399$ eV (schematic in Fig. 4C) is well agreement with the *ab initio* prediction of ~0.40 eV and the 0.385 ± 0.143 eV inferred from the reported data of doped a-Si:H (*36*). Fig. 4D finds $E_a$ of doped a-Si:H is noticeably smaller than that of the intrinsic counterpart (*34, 36-39*), most likely owning to existence of the exclusive metastable H configurations in doped materials (inferred from Fig. 2A). We notice phosphorus doped a-Si:H also has smaller $E_a$, thus it is expected to make similar contributions to the *abnormal* Staebler-Wronski effect. This speculation is evident from the light-soaking behavior of "half" cells with structure of Ag/IWO/*n*-a-Si:H/*i*-a-Si:H/*n*-c-Si/IWO/Ag,

where the *p*-a-Si:H is totally removed (Fig. S11). Given doped a-Si:H has small $E_a$ but great $\tau_{WW}$ at low temperatures, we conclude that the cold climates can effectively prevent the decay of metastable FF.

In summary, it is the first time to observe the *abnormal* Staebler-Wronski effect in doped a-Si:H "thin" films, which is appealing for realizing outstanding optoelectronic devices. Specifically, we demonstrate this *abnormal* effect noticeably improves the charge carrier transport in SHJ solar cells, yielding a FF of 85.42%. Future research should be focused on implementing this effect to perovskite/SHJ tandem solar cells, aiming to approach its theoretical FF constrained by the Schokley-Queisser limit (*40*, *41*).

**ACKNOWLEDGEMENTS**

W. L. acknowledges Ms. Jiajia Ling, Prof. Zhongquan Ma and Dr. Wenbo Ji for their fruitful discussions. Authors acknowledge CIC Choshu Industry Co. Ltd (Japan) for their stability measurements. **Funding:** This work is supported by two National Natural Science Foundations of China (No. 62004208 and No. 62074153) and Shanghai Scientific Innovation Foundation (No. 19DZ1207602). **Author contributions:** W. L. conceived the idea, designed the overall experiments, proposed the *abnormal* Staebler-Wronski effect and led the project. J. S., L. Z., W. L., A. H., F. M., Y. X., H. L. performed the device optimization. W. L. performed molecular dynamics and *ab initio* simulations, and modeled the experimental data. W. L. performed the SIMS analysis. A. H. performed the I-V measurements under 11 suns, 48 suns and 60 suns. S. H., X. L., J. P., Y. Y., K. J. and Z. M. measured the 1-sun FF evolutions. J. D. performed the RCA cleaning. Y. Yu performed the system calibrations of the solar simulator. Z. Li performed the FTIR measurements. H. Z. and J. S. performed the FF evolutions at different temperatures. Y. Y., K. J. and S. H. performed the Sinton measurements. Y. Y. and W. L. performed the dark I-V measurements. X. L. performed the damp-heat measurements. Y. Y., K. J., X. L. and Y. Yao performed the pFF measurements. Z. D., J. Y., X. Y., X. S., L. X. and J. K. participated in the discussion of molecular dynamics simulations. Y. G. and F. L. performed the TA measurements and analysis. W. L. and Z. L. supervised the project. W. L. wrote the paper. All authors contributed to the discussion of the results and revision of the manuscript. **Competing interests:** The authors declare no competing interests. **Data and materials availability:** All data regarding this work are provided in the main manuscript and the supplementary materials.

**SUPPLEMENTARY MATERIALS**

Materials and Methods

Figs. S1 to S11

Tables S1 and S2

References (*42–44*)

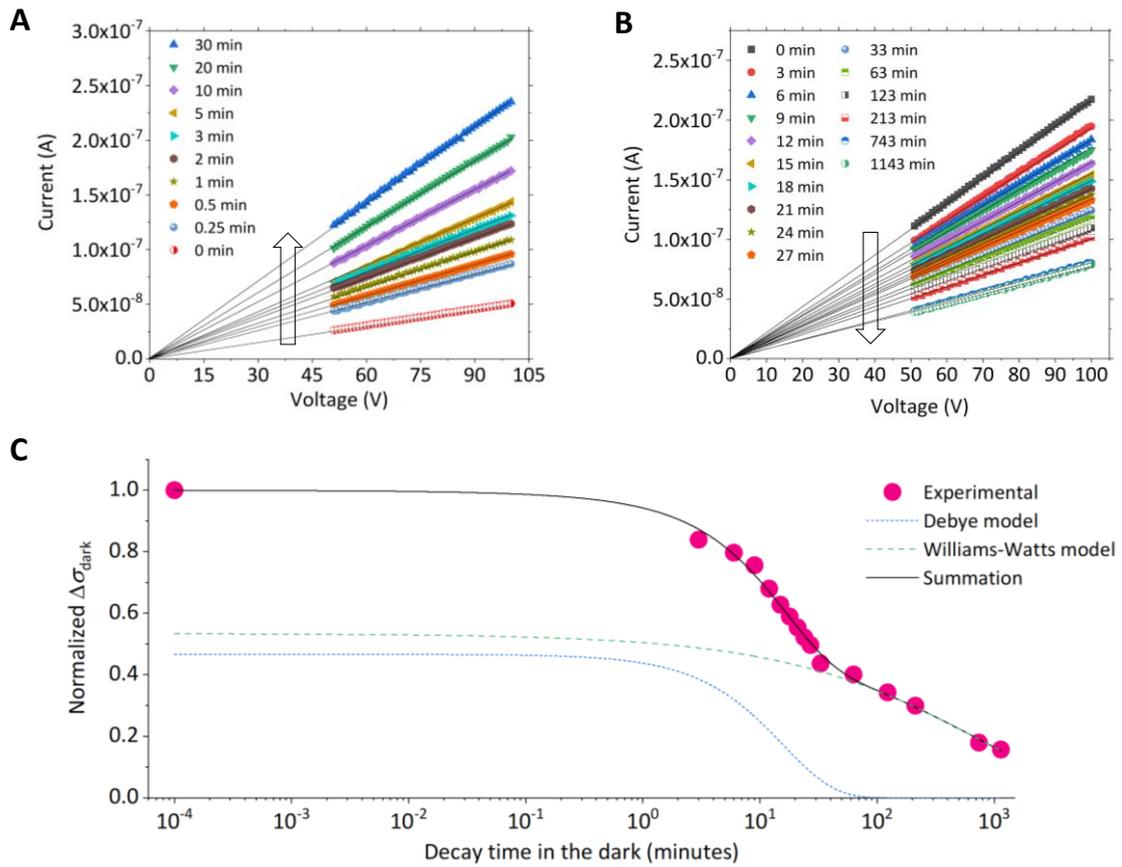

**Fig. 1. Discovery of the *abnormal* Staebler-Wronski effect.** (**A**) Dark current-voltage evolution of *p*-a-Si:H thin film as a function of light-soaking time under one-sun illumination. (**B**) Dark current-voltage evolution of light soaked *p*-a-Si:H thin film as a function of "sleeping" time in the dark. (**C**) Normalized decay of $\Delta\sigma_{dark}$ fits to a combination of Debye model and Williams-Watts model.

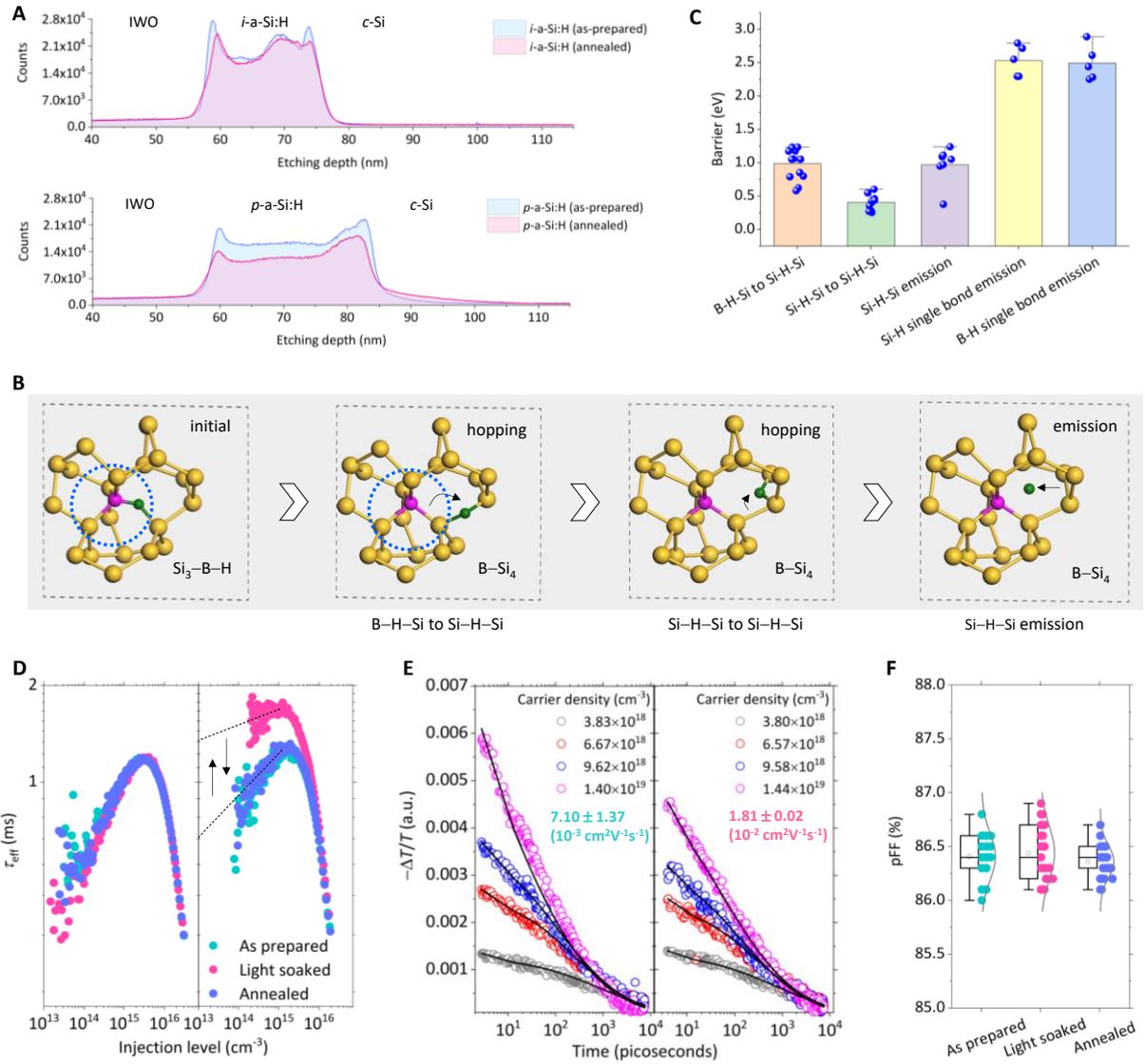

**Fig. 2. Mechanisms of the *abnormal* Staebler-Wronski effect.** (**A**) H⁻ profiles of *i*-a-Si:H and *p*-a-Si:H before and after 30-minute annealing at 180 ℃. (**B**) Schematic H movements in *ab initio* dynamics, the yellow, magenta and cyan balls respectively represent Si, B and H atoms. (**C**) Energy barriers of H movements. (**D**) Injection-dependent $\tau_{\mathrm{eff}}$ of *p*-a-Si:H/*i*-a-Si:H/*n*-c-Si/*i*-a-Si:H/*p*-a-Si:H (right) and *i*-a-Si:H/*n*-c-Si/*i*-a-Si:H (left) before and after two-hour light soaking under one-sun illumination, as well as that after 15-minute annealing at 180 ℃ (before which they are light soaked). (**E**) TA signals of *p*-a-Si:H before (left) and after (right) 2-hour light soaking under one-sun illumination, the fitting is based on the one-dimension recombination and diffusion model. (**F**) pFF of complete SHJ solar cells before and after two-hour light soaking under one-sun illumination, as well as that after 15-minute annealing at 180 ℃ (before which they are light soaked).

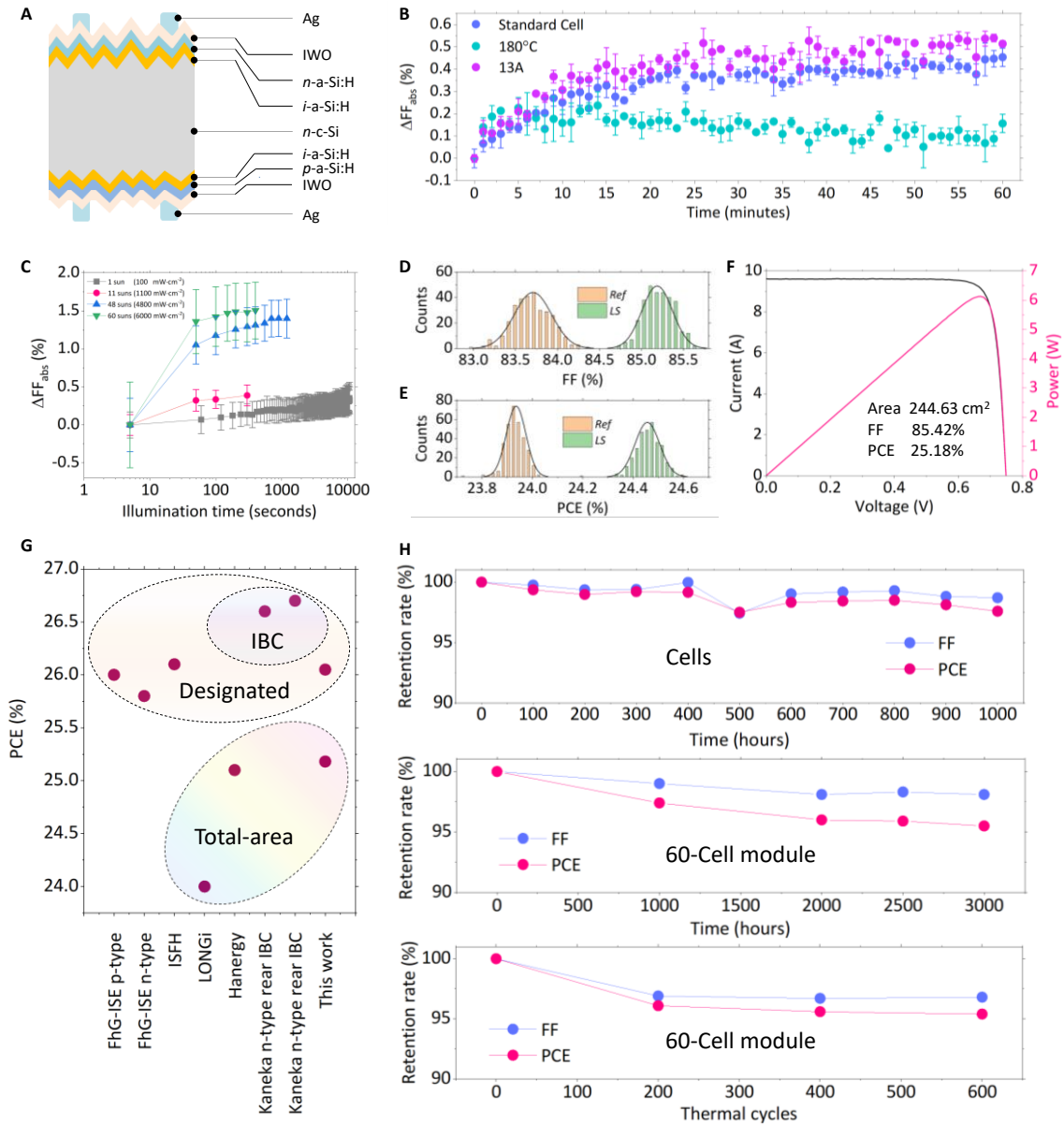

**Fig. 3. Improvement of SHJ solar cells by the *abnormal* Staebler-Wronski effect.** (**A**) Device structure in this work. (**B**) FF evolution of devices in "Standard Cell" and "180 °C" measured as a function of light-soaking time under one-sun illumination; cells in "Standard Cell" are as prepared, while cells in "180 °C" are annealed for two hours at 180 °C. FF evolution of devices in "13A" measured as a function of electrifying time at 13 amperes. (**C**) FF evolution under different light intensities. (**D**) FF and (**E**) PCE of SHJ solar cells before (*Ref*) and after (*LS*) 70-second light soaking under 60-sun illumination. (**F**) Certificated current-voltage and power-voltage curves of a 244.63-cm² SHJ solar cell. (**G**) Comparison of PCE in this work with the best c-Si solar cells, where the "designated-area" PCE is measured on our own CNAS (China National Accreditation Service)

platform, while the "total-area" PCE is independently certificated by National Photovoltaic Industry Metrology and Testing Center (NPVM). (**H**) 1000-hour degradations of FF and PCE of 6-inch devices at DH85, 3000-hour degradations of FF and PCE of a 60-cell module at DH85, and degradations of FF and PCE of a 60-cell module during 600 thermal cycles between -40 ℃ and 85 ℃. All measurements are conducted under the standard conditions (25 ℃, 100 mW cm$^{-2}$).

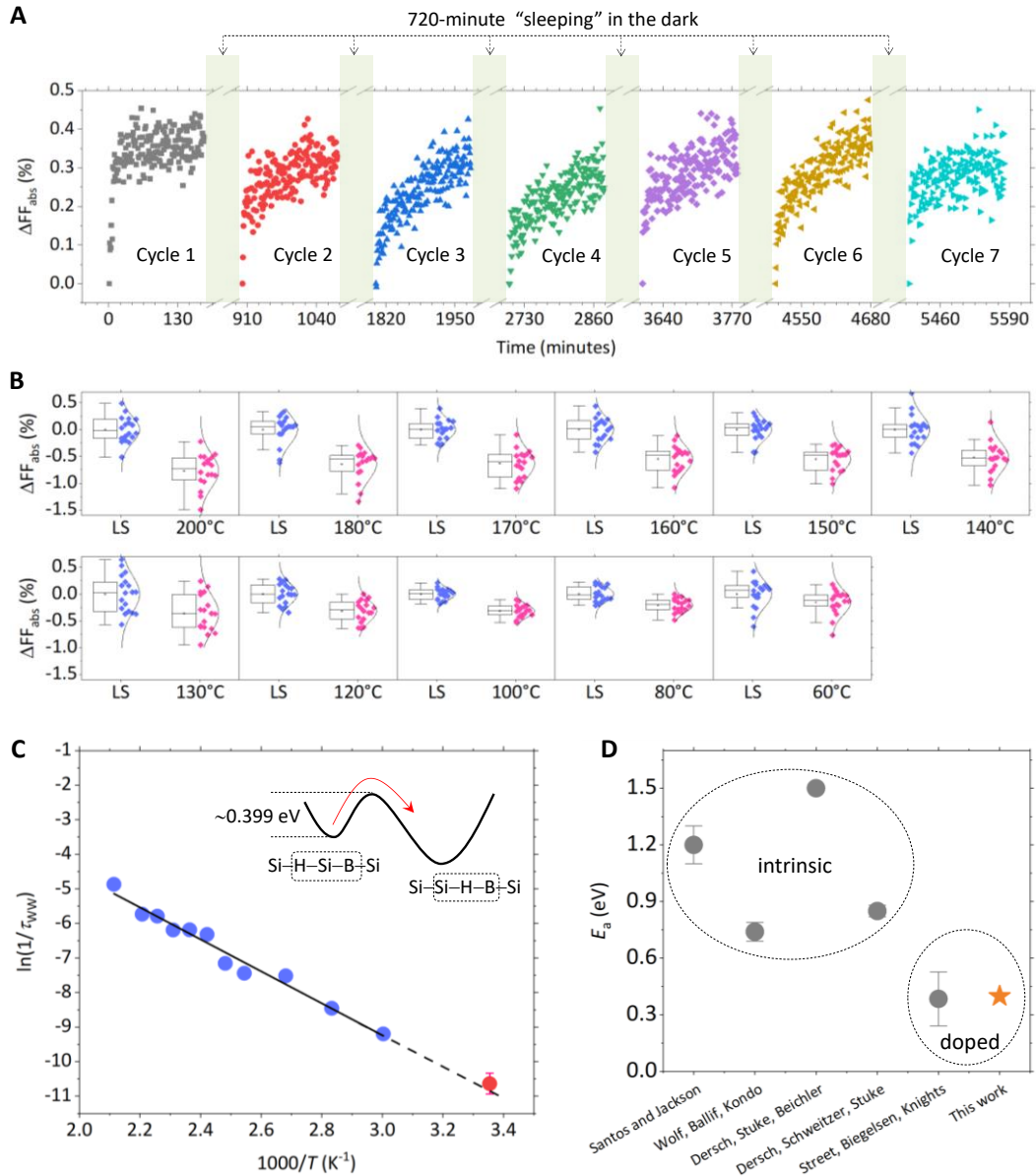

**Fig. 4. Reversible behavoir of the *abnormal* Staebler-Wronski effect.** (**A**) Evolutions of FF (compared to as-prepared state) in cycles composed of alternant 180-minute one-sun illumination and 720-nimute "sleeping" in the dark. (**B**) Decay of FF in 10 minutes at different temperatures, the "LS" states are light soaked under 60-sun illumination for 70 seconds, followed by 25-minute dark "sleeping". (**C**) Fitting of equation (3) to the $\tau_{WW}$ (blue circles) derived from experimental average ΔFF in Fig. 4b, the red circle is picked from Table S1 measured at 25 °C. The inset is schematic of the H hopping barrier, where the 0.399 eV is calculated from the slope of the fitting line. (**D**) Typical $E_a$ of intrinsic and doped a-Si:H materials.

# Supplementary Materials for

**Abnormal Staebler-Wronski effect of amorphous silicon**

Wenzhu Liu*[†], Jianhua Shi*, Liping Zhang*, Anjun Han*, Shenglei Huang, Xiaodong Li, Jun Peng, Yuhao Yang, Yajun Gao, Jian Yu, Kai Jiang, Xinbo Yang, Zhenfei Li, Junlin Du, Xin Song, Youlin Yu, Zhixin Ma, Yubo Yao, Haichuan Zhang, Lujia Xu, Jingxuan Kang, Yi Xie, Hanyuan Liu, Fanying Meng, Frédéric Laquai, Zengfeng Di, Zhengxin Liu[†]

Correspondence to: wenzhu.liu@mail.sim.ac.cn (W.L.); z.x.liu@mail.sim.ac.cn (Z.L.)

**This PDF file includes:**

    Materials and Methods
    Figs. S1 to S11
    Tables S1 and S2
    References (*42–44*)

**Materials and Methods**

Material characterization

Glow-discharge $p$-a-Si:H films with thickness of ~60 nm are deposited onto silica glasses in 40.68-MHz very-high-frequency plasma-enhanced chemical vapor deposition (VHF-PECVD; IE Sunflower, OAK-DU-5) at 200 °C. Then 200-nm-thick 10 mm × 2 mm silver strips are thermally evaporated on the film surface to form the transmission-line-model structure (*42*). Current-voltage characteristics between the two strips with interval distance of 200 μm are probed by KEITHLEY 6487. For Fourier transform infrared spectroscopy (FTIR; Perkin Elmer, Spectrum 100) measurements, $p$-a-Si:H thin films are deposited on >3000 Ω cm float-zone c-Si substrates, their infrared absorptions from Si–H and B–H bonds are characterized using the transmission mode. H$^-$ profiles of IWO/$i$-a-Si:H/c-Si, IWO/$p$-a-Si:H/c-Si and $p$-a-Si:H/$i$-a-Si:H/c-Si are analyzed by ToF-SIMS (ION TOF, GmbH-Muenster, Germany), during which the chamber pressure, primary ion source and current are $1.0 \times 10^{-9}$ mbar, 30-keV Bi$^+$ and 1.0 pA respectively, the depth profiles are acquired using 500-eV Cs$^+$ sputter beam. Cross-sectional images of $p$-a-Si:H are probed by high-resolution transmission electron microscope (FEI Titan 80-300ST), operated at 200 kV. Injection-dependent $\tau_{\text{eff}}$ and pFF are measured by the Sinton WCT-120 and Suns-Voc respectively. Ultrafast and broadband TA spectra are carried out using a homebuilt pump–probe setup. The output of a titanium sapphire amplifier (Coherent LEGEND DUO, 4.5 mJ, 3 kHz, 100 fs) splits into three beams (2.0 mJ, 1.0 mJ and 1.5 mJ), two of which separately pump two optical parametric amplifiers (OPA; Light Conversion TOPAS Prime). TOPAS-1 provides tunable pump pulses, and TOPAS-2 generates the probe pulses. A 1300 nm pulse from TOPAS-2 is sent through a calcium fluoride (CaF2) crystal mounted on a continuously moving stage. This generates a white-light supercontinuum pulses from 350 nm to 1100 nm. The pump pathway length is varied between 5.12 m and 2.60 m with a broadband retroreflector mounted on an automated mechanical delay stage (Newport linear stage IMS600CCHA controlled by a Newport XPS motion controller), thereby generates delays between pump and probe from −400 picoseconds to 8 nanoseconds. Pump and probe beams are overlapped on surface of the $p$-a-Si:H. By a beam viewer (Coherent, LaserCam-HR II) we regulate the size of pump beam about three times larger than the probe beam. The probe beam is guided to a custom-made prism spectrograph (Entwicklungsbüro Stresing) where it is dispersed by a prism onto a 512 pixel complementary metal-oxide semiconductor (CMOS) linear image sensor (Hamamatsu G11608- 512DA). The probe pulse repetition rate is 3 kHz, while the excitation pulses are mechanically chopped to 1.5 kHz (100 fs to 8 ns delays), while the detector array is read out at 3 kHz.

Molecular dynamics and *ab initio* dynamics

The dynamics simulations are conducted on the platform of Materials Studio 2017R. We use the *Tersoff* empirical potential to express the atomic interactions for a-Si and a-Si:H, which has been demonstrated for amorphous tetrahedral semiconductors (*43*). The c-Si supercells are heated up to 3000 K with a rate of 540 K/ps (picoseconds), followed by 200-ps equilibrium in order to fully destroy small crystallinities. Then the metallized liquid silicon is quenched to 300 K in a slow rate of $1.0 \times 10^{11}$ K/s. The slow cooling efficiently reduces structure defects, such as dangling bonds and floating bonds. To gain the *p*-a-Si:H model, we replace some four-coordinated Si atoms by B atoms, and run *ab initio* dynamics for 2 ps to relax its geometry. To investigate the capture of H by B, we place H atoms >2.3Å away (much longer than B–H and Si–H single bonds) from the B atoms, and run the *ab initio* dynamics again to observe the coordinate changes. The functional, ensemble, temperature and simulation time are GGA-PBE, NVT, 298 K and 0.3 ps, respectively. The reaction barriers are calculated by transition-state searches between optimized structures using the generalized synchronous transit method (*44*).

Device fabrication

Czochralski (CZ) *n*-c-Si wafers are purchased from Sichuan Yongxiang Silicon Material Co. Ltd. Their initial thickness and electrical resistivity are ~160 μm and 0.3–2.1 Ω cm respectively. The saw damage is etched in 20.0 vol% alkaline solution at 80 °C for 2 minutes, followed by formation of surface pyramids via immersion in 2.1 vol% alkali solution at 80 °C for 8 minutes. Then, they experience standard RCA cleaning to remove surface organics and metal ions. After that, these wafers undergo 3-minute dipping in 2.0 % hydrofluoric acid water solution to remove the surface oxide. In the chambers of VHF-PECVD, 5-nm *i*-a-Si:H, 15-nm *p*-a-Si:H, 4-nm *i*-a-Si:H and 6-nm *n*-a-Si:H are sequentially deposited on the two faces of the clean wafers. The process temperatures are 200 ± 10 °C. The *i*-a-Si:H layer consists of two sub-layers, their power density and chamber pressure during deposition are 67/40 mW cm$^{-2}$ and 50/80 Pa respectively. The first layer is deposited using pure SiH$_4$, while the second layer is deposited using diluted SiH$_4$ in H$_2$ with a ratio of [SiH$_4$]:[H$_2$] = 1:10. 15-second H$_2$ plasma is applied to treat the two *i*-a-Si:H films for improving the passivation quality at the *i*-a-Si:H/*n*-c-Si interfaces. Power density, chamber pressure and gas flow ratio during deposition of the *n*-a-Si:H are 33 mW cm$^{-2}$, 80 Pa and [PH$_3$]:[SiH$_4$]:[H$_2$] = 1.5:100:1000. The *p*-a-Si:H layer also has two sub-layers, whose power density, chamber pressure and gas flow ratio during deposition are 20/20 mW cm$^{-2}$, 80/80 Pa and [B$_2$H$_6$]:[SiH$_4$]:[H$_2$] = 1:100:100/2:100:400 respectively. Tungsten-doped indium oxide (IWO) is grown by the reactive plasma deposition at 150 ℃, whose target material is 1.0 % tungsten doped in indium oxide. Silver busbars and fingers are screen printed on the two faces of the devices using low-temperature paste, followed by annealing at 150 ℃ for 5 minutes and 185 ℃ for 30 minutes. For the certificate cell, an 80-nm SiO$_x$ layer is capped onto the sun-side surface in a 13.56 MHz radio-frequency PECVD (ULVAC CME-400).

Device characterization

Current-voltage characteristics of all solar cells without SiO$_x$ antireflection are tested under standard conditions (25 ℃, 100 mW cm$^{-2}$) in a solar simulator (Halm IV, ceitsPV-CTL2). The light intensity is calibrated using a certified NREL (National Renewable Energy Laboratory) reference cell. The submitted cell with SiO$_x$ antireflection is independently tested by NPVM (National Photovoltaic Industry Metrology and Testing Center) in Fujian Province, China, one of the designated test centers for the *solar cell efficiency tables*. The device area is captured by an automatic image test system. Before the certification, it is light soaked for 30 minutes under one-sun illumination, followed by cooling down to room temperature. The conveyor during light soaking is pre-heated to ~200 ℃, and the light intensity is adjusted from one sun to 60 suns (JBAO Technology Ltd, NLIDR-S60). These cells are quickly cooled down by cold-air blowing after the light soaking. All devices are measured under standard conditions.

Damp-heat degradation

The devices undergo 1000-hour damp-heat impact at DH85 in the dark, during which they are in open-circuit condition. These devices are 6-inch SHJ solar cells without any encapsulations. The 60-cell module undergoes 3000-hour damp-heat impact at DH85 in the dark according to the IEC 60068-2-78, during which it is in open-circuit condition. All measurements are conducted under standard conditions (25 ℃, 100 mW cm$^{-2}$), out of the climate test chamber.

Thermal cycle degradation

The thermal cycles are conducted in accordance with the IEC 61215-2:2016 at CIC Choshu Industry Co. Ltd, Japan. The cycle temperature is between -40 ℃ and 85 ℃. The applied current is 100% $I_{mpp}$ at the rising edge of temperature.

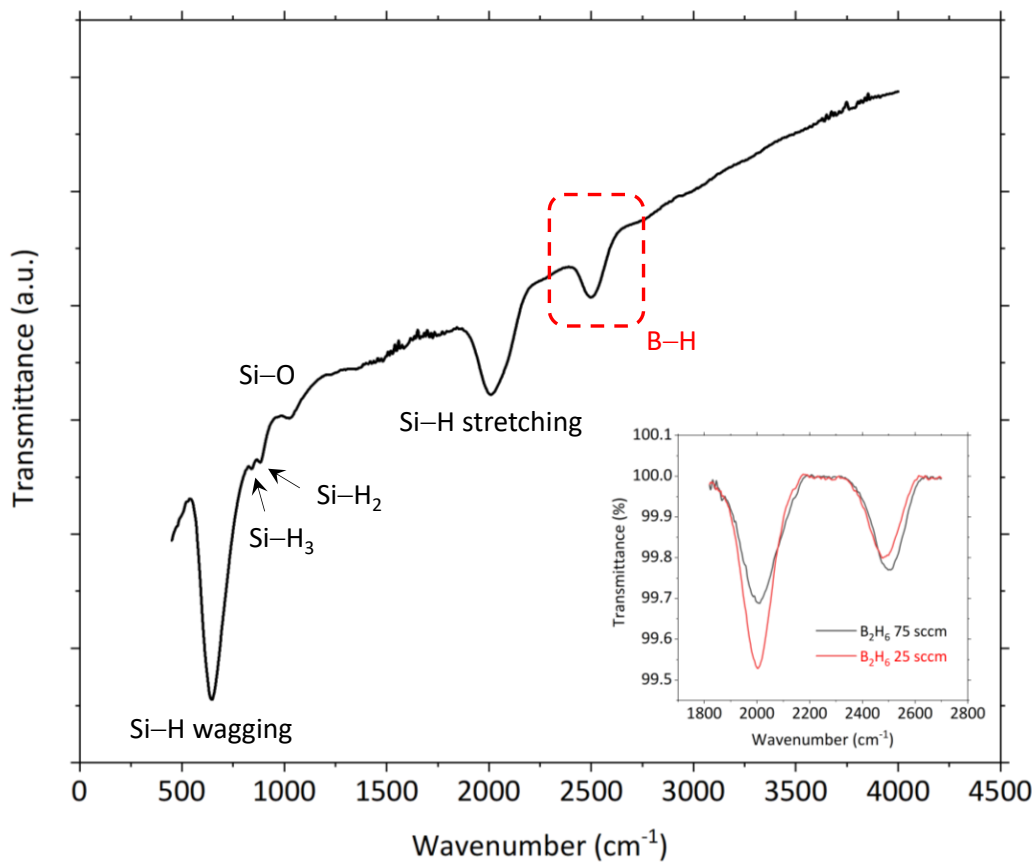

**Fig. S1. FTIR spectrum of a *p*-a-Si:H thin film.** The absorption model at ~2500 cm$^{-1}$ originates from B−H bonds (*1*). Inset is FTIR spectra of two *p*-a-Si:H films, whose flow rates of $B_2H_6$ during deposition are 75 standard cubic centimeter per minute (sccm) and 25 sccm respectively, other parameters are kept the same as each other. Evidently, higher $B_2H_6$ flow rate increases the B−H intensity but decreases the Si−H intensity. This indicates H atoms prefer to be captured by surrounding B atoms in comparison to Si atoms.

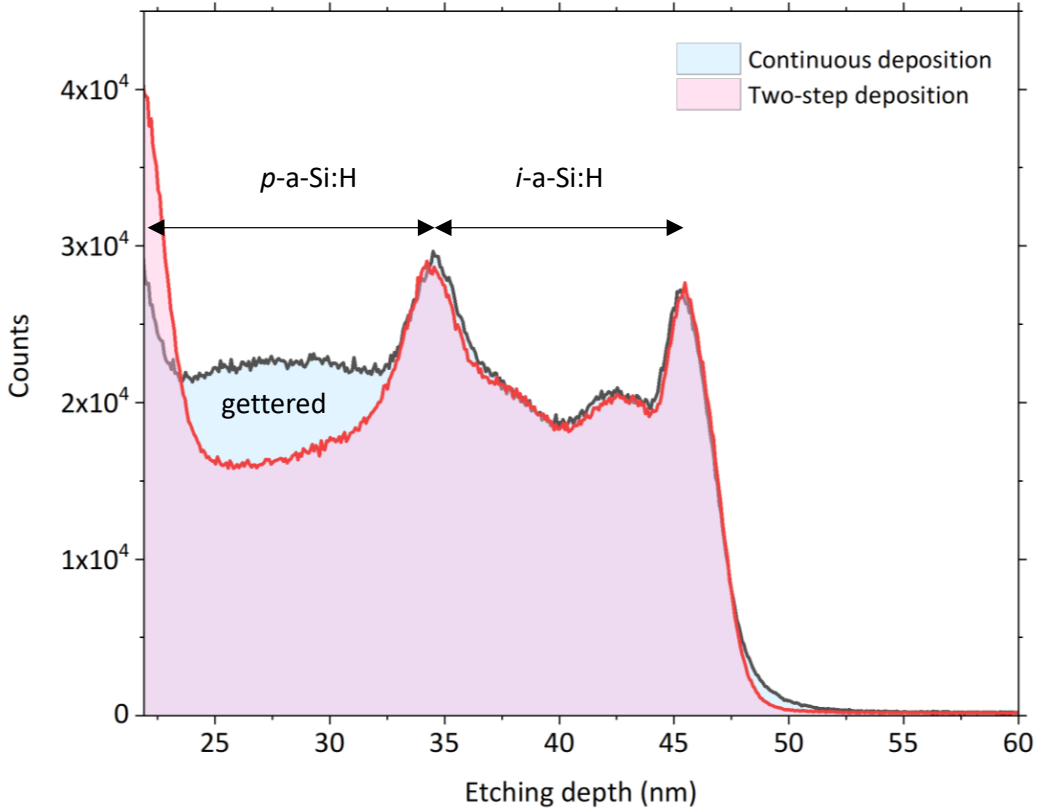

**Fig. S2. H⁻ spectra of *i*/*p*-a-Si:H stack probed by ToF-SIMS.** The two samples have the same structure of *i*/*p*/*i*/*p*-a-Si:H. For "Continuous deposition", all layers of *i*/*p*/*i*/*p*-a-Si:H are continuously deposited on a c-Si substrate in PECVD chambers without air exposure in between. As a reference, the first *i*/*p*-a-Si:H stack of "Two-step deposition" is deposited on the c-Si substrate, followed by slow oxidation in $N_2$ cabinet for two days, after then the second *i*/*p*-a-Si:H stack is capped onto the first *i*/*p*-a-Si:H stack. To be clear, only the first *i*/*p*-a-Si:H stack contacting with the c-Si substrate is shown. It is evident that the slow oxidation getters huge amounts of H atoms in the *p*-a-Si:H from inside to surface, whereas the H content in the *i*-a-Si:H almost remains unchanged.

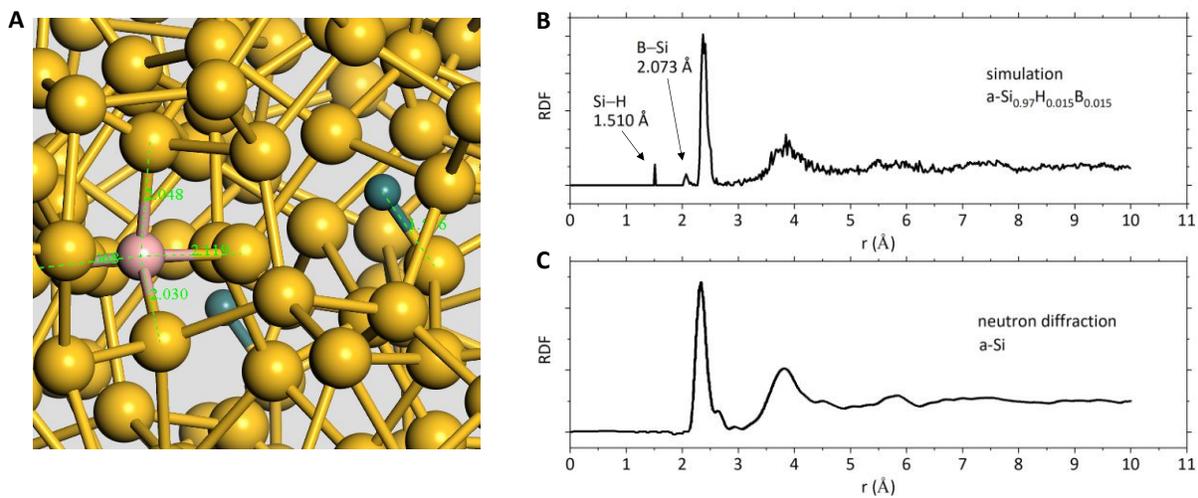

**Fig. S3. RDF of a-Si$_{0.97}$H$_{0.015}$B$_{0.015}$.** (**A**) Local structure of a *p*-a-Si:H. The yellow, magenta and cyan balls represent Si, B and H respectively. (**B**) Simulated RDF of the *p*-a-Si:H, in which the small peaks at 1.510Å and 2.073Å respectively stem from Si–H and B–Si bonds. (**C**) RDF of a pure a-Si measured by neutron diffraction (*2*).

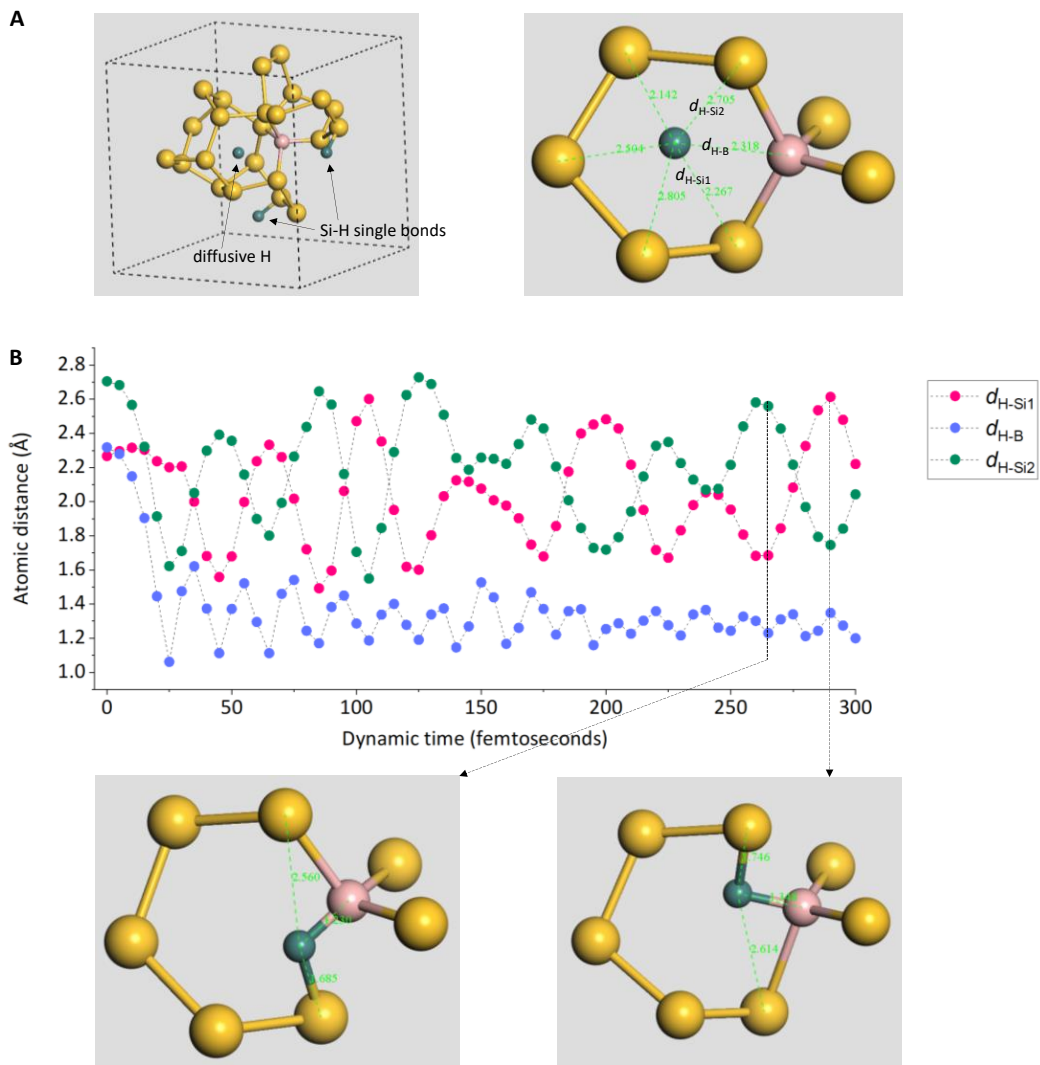

**Fig. S4. *Ab initio* dynamics of a diffusive H when passes by a B−Si$_4$ site.** (**A**) Illustration of the diffusive H, initially, its distances from all adjacent atoms are >2.14 Å. The yellow, magenta and cyan balls respectively represent Si, B and H atoms. (**B**) Dynamic evolutions of the three characteristic distances, $d_{H-Si1}$, $d_{H-B}$ and $d_{H-Si2}$ in a NVT ensemble at 298 K. Evidently, the B rapidly captures the H, judged from the fact that $d_{H-B}$ decreases from 2.32 Å to 1.28 ± 0.05 Å within 200 femtoseconds. In addition, it is also noted that $d_{H-Si1}$ and $d_{H-Si2}$ have opposite phases, indicating the H switches between the adjacent B−H−Si bridged sites (as schematic by the two local structures).

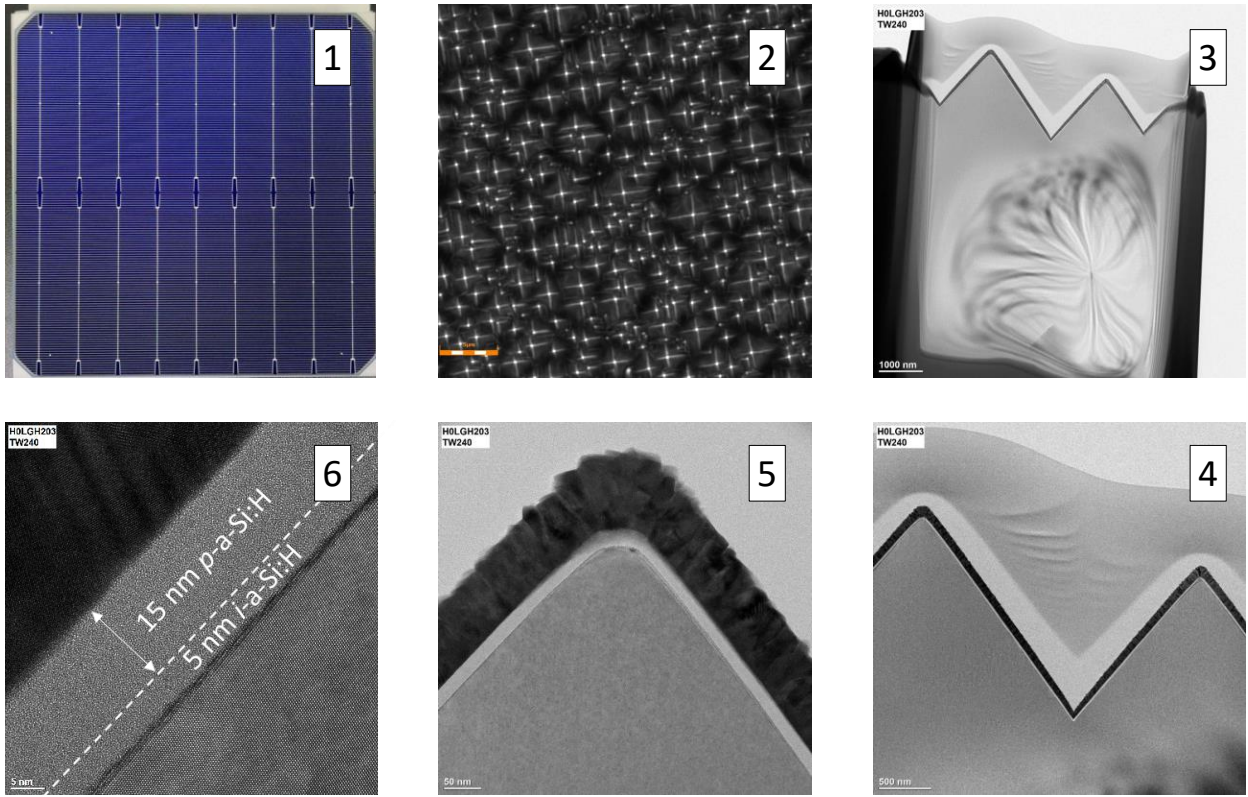

**Fig. S5**. **Thickness of *p*-a-Si:H in the SHJ solar cell.** High-resolution transmission electron microscope images show that total thickness of *i*/*p*-a-Si:H on the pyramids is ~20 nm. In combination with ellipsometric analysis, thickness of the *p*-a-Si:H is distinguished to be ~15 nm. The interfaces show conformal contacts at *i*-a-Si:H/c-Si and *p*-a-Si:H/IWO.

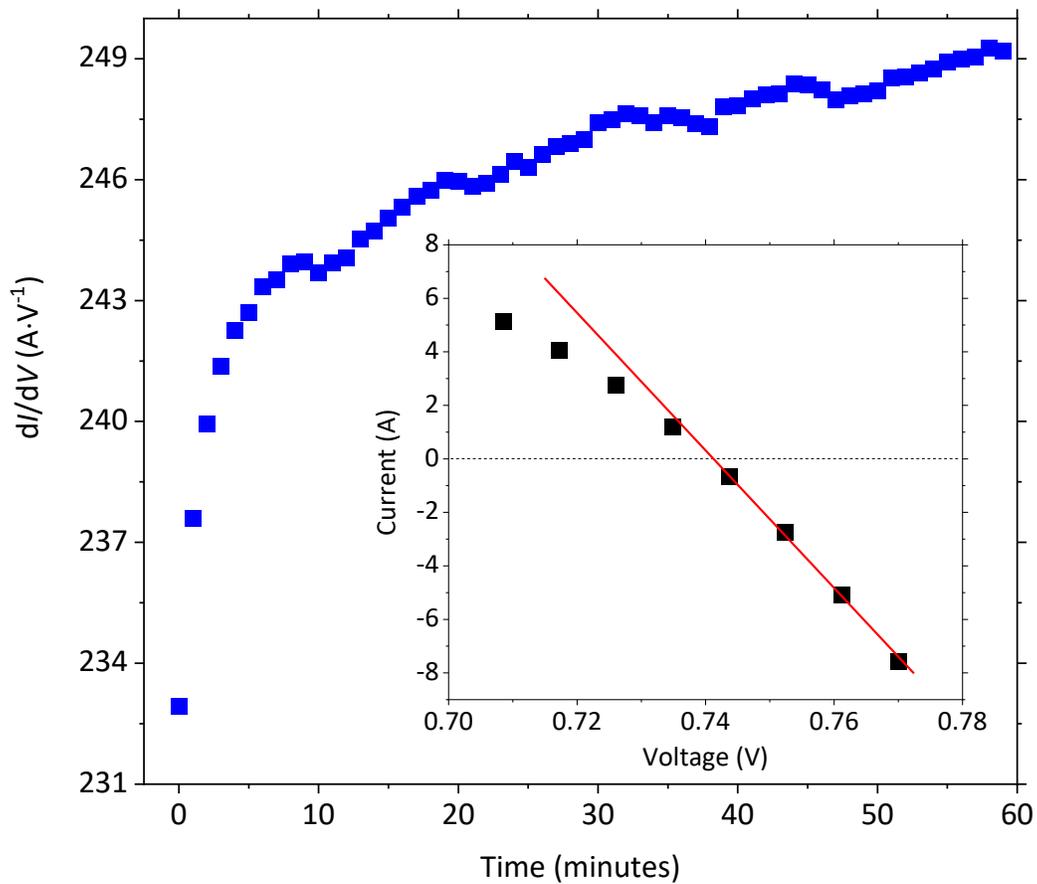

**Fig. S6. Slope of the light current-voltage curve near the low-internal-field region.** The d$I$/d$V$ near the low-internal-field region plots as a function of the light-soaking time under one-sun illumination. The inset shows a typical fitting.

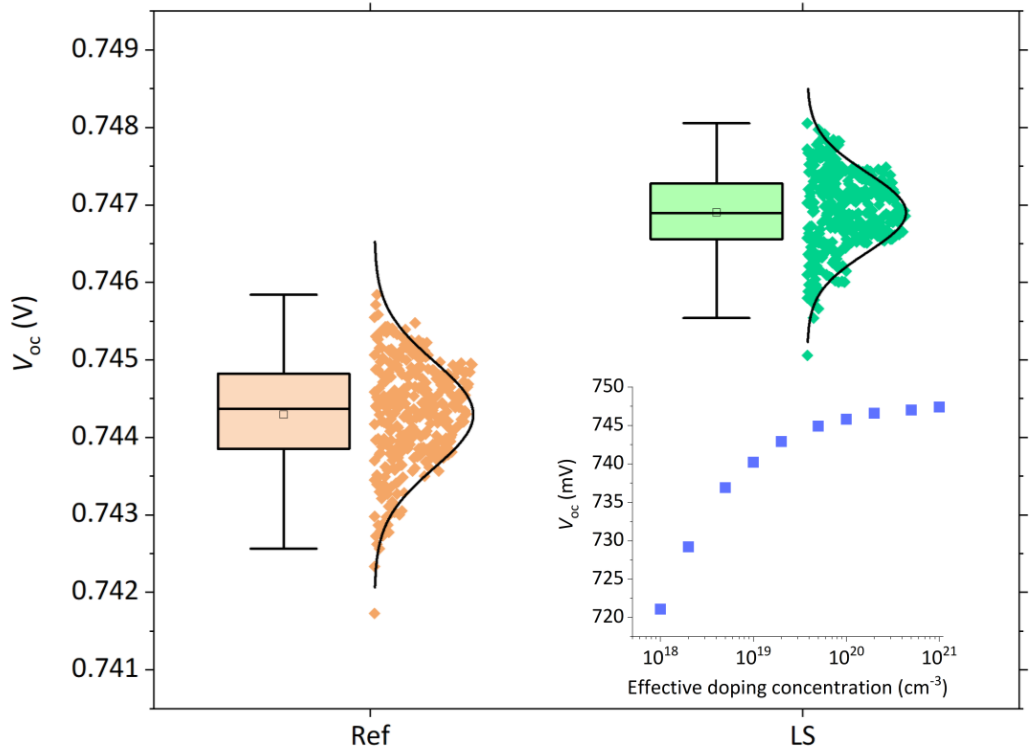

**Fig. S7**. **Improvement of $V_{oc}$.** The average $V_{oc}$ increases from 0.7443 V (Ref) to 0.7469 V (LS) after 70-second light soaking under 60-sun illumination. Inset is the theoretical $V_{oc}$ as a function of effective doping concentration of B atoms in the *p*-a-Si:H "thin" film. Detailed simulation parameters are provided in Table S2.

**Fig. S8. Certificate report.** Certificated results from National Photovoltaic Industry Metrology and Testing Center (NPVM), one of the designated test centers for *solar cell efficiency tables* (*3*).

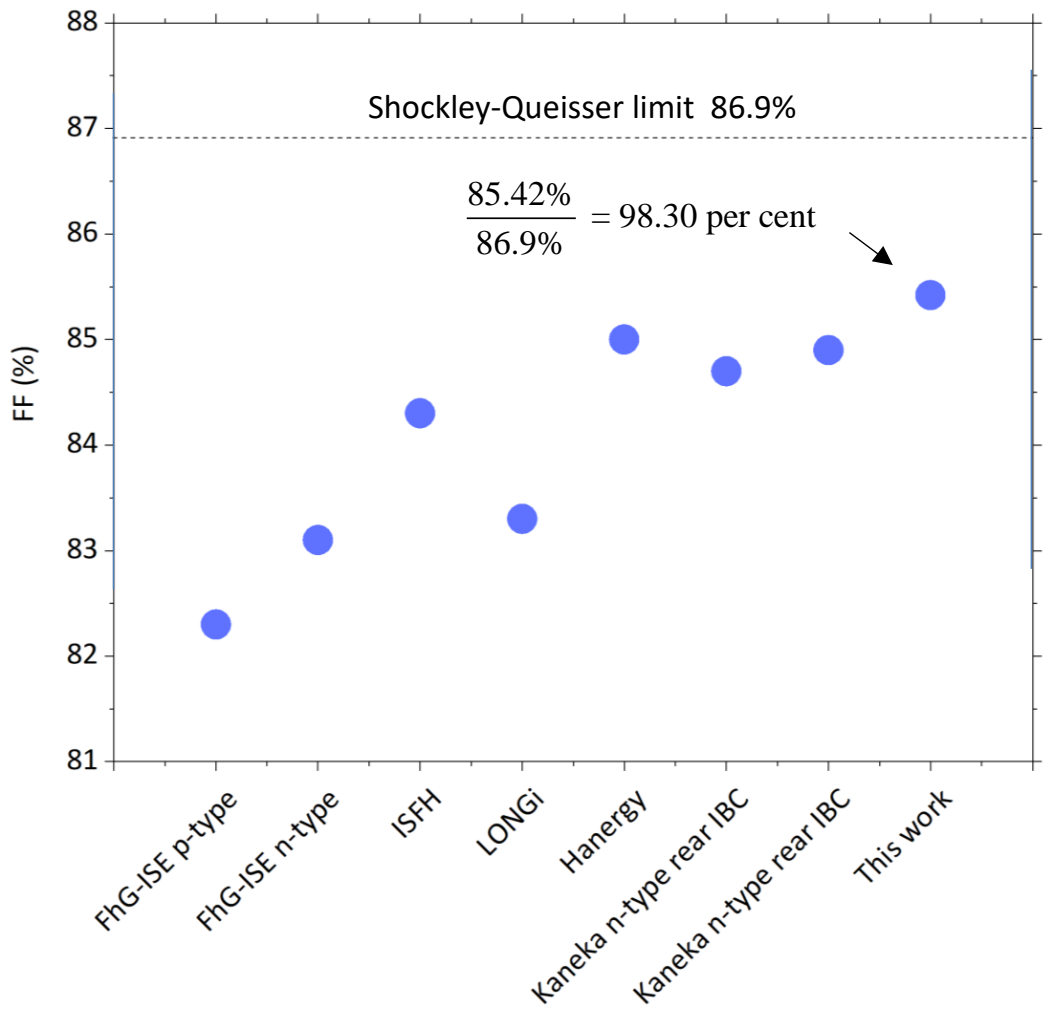

**Fig. S9. FF of the best crystalline silicon solar cells.** Devices in the color region are measured on total area.

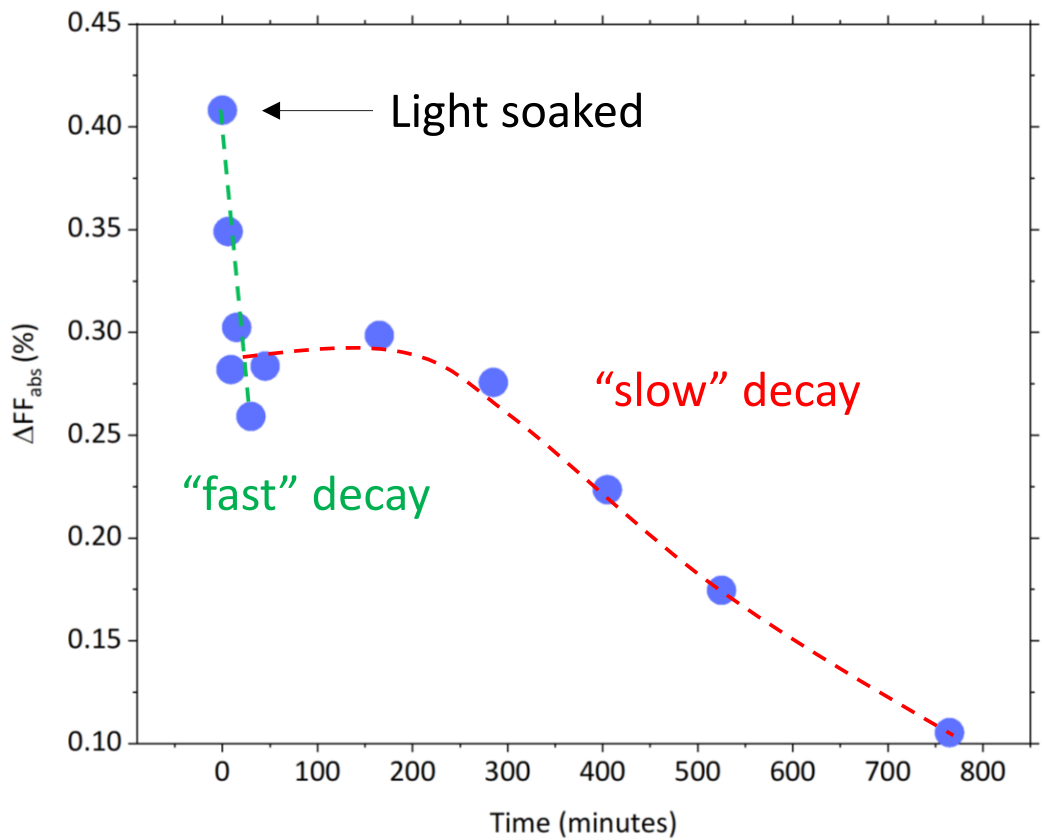

**Fig. S10. Evolution of FF in the dark.** The cell is light soaked under one sun for 180 minutes, after then it "sleeps" in the dark for 765 minutes. The FF is monitored for 11 times during this period.

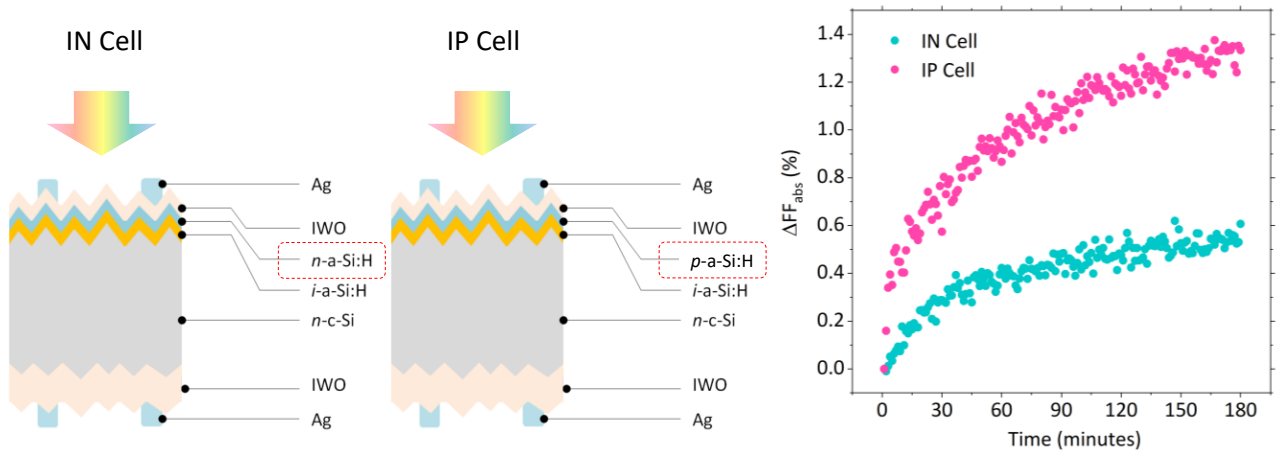

**Fig. S11. FF evolution of two "half" SHJ solar cells under one-sun illumination.** "IN Cell" is a high-low-junction cell without *p*-a-Si:H, while "IP Cell" is a *p-n*-junction cell without *n*-a-Si:H. Similar FF behaviors are observed. The higher ΔFF of the "IP Cell" over that of the "IN Cell" indicates the *abnormal* Staebler-Wronski effect is dominated by the *p*-a-Si:H. (Notes: The higher ΔFF of the "IP Cell" over that of the standard SHJ solar cell is likely due to its low initial value caused by poor passivation).

**Table S1.** Fitting parameters of equation (1) in Fig. 1C, and $\beta_{WW}$ from another study (*35*).

| $\Delta\sigma_D$ | $\Delta\sigma_{WW}$ | $\tau_D$ (minutes) | $\tau_{WW}$ (minutes) | $\beta_D$ | $\beta_{WW}$ This work | $\beta_{WW}$ Kakalios |
|---|---|---|---|---|---|---|
| 0.466 | 0.534 | 15.84 | 694.43 | 1 | 0.44 | 0.45 |
| (0.047) | | (1.55) | (181.37) | | (0.07) | (0.05) |

Constraint condition: $\Delta\sigma_D + \Delta\sigma_{WW} = 1$ and $\beta_1 = 1$.

**Table S2.** Parameters for simulation of SHJ solar cells.

|  | $i$-a-Si:H[a] | $n$-a-Si:H | $i$-a-Si:H[b] | $p$-a-Si:H | $n$-c-Si |
|---|---|---|---|---|---|
| Thickness (nm) | 5 | 5 | 4 | 15 | 120000 |
| Dielectric constant | 11.9 | 11.9 | 11.9 | 11.9 | 11.9 |
| Electron affinity (eV) | 3.8 | 3.8 | 3.8 | 3.8 | 4.05 |
| Bandgap (eV) | 1.74 | 1.74 | 1.74 | 1.70 | 1.12 |
| Effective conductive band density (cm$^{-3}$) | $1.0\times10^{20}$ | $1.0\times10^{20}$ | $1.0\times10^{20}$ | $1.0\times10^{20}$ | $2.8\times10^{19}$ |
| Effective valence band density (cm$^{-3}$) | $1.0\times10^{20}$ | $1.0\times10^{20}$ | $1.0\times10^{20}$ | $1.0\times10^{20}$ | $2.7\times10^{19}$ |
| Electron mobility (cm$^2$·V$^{-1}$ s$^{-1}$) | 20 | 10 | 20 | 5 | 1321 |
| Hole mobility (cm$^2$·V$^{-1}$ s$^{-1}$) | 2 | 1 | 2 | 1 | 461 |
| Acceptor concentration (cm$^{-3}$) | 0 | 0 | 0 | *variable* | 0 |
| Donor concentration (cm$^{-3}$) | 0 | $1.0\times10^{19}$ | 0 | 0 | $2.0\times10^{15}$ |

Notes: [a]Sunside $i$-a-Si:H passivation layer, [b]rear $i$-a-Si:H passivation layer. A defect interface with a density of $2.0\times10^{10}$ cm$^{-2}$ is placed between $n$-c-Si and $i$-a-Si:H on the rear side.